\documentclass[a4paper,11pt]{article}
\usepackage{jinstpub} 
\usepackage{lineno}
\usepackage{upgreek}
\usepackage{siunitx}

\proceeding{21$^{\text{st}}$ ``Trento'' Workshop on Advanced Silicon Radiation Detectors\\
February 17--19, 2026\\
Perugia, Italy}

\usepackage{graphicx}
\usepackage{subcaption}   

\title{\boldmath Measurements and simulations of X-ray radiation damage effects on
CNM n-type 4H-SiC MOS capacitors}

\author[a,1]{K. Aouadj,\note{Corresponding author.}}
\author[b,a]{A. Fondacci,}
\author[a]{A. Morozzi,}
\author[c,a]{D. Passeri,}
\author[a,d]{T. Croci,}
\author[e]{S. A. Onder,}
\author[e]{D. Radmanovac,}
\author[e]{T. Bergauer,}
\author[d,f]{S. Mattiazzo,}
\author[g,a]{and F. Moscatelli}

\affiliation[a]{INFN Perugia, via A. Pascoli, Perugia, Italy}
\affiliation[b]{Department of Physics, University of Perugia, via A. Pascoli, Perugia, Italy}
\affiliation[c]{Department of Engineering, University of Perugia, via G. Duranti 93, Perugia, Italy}
\affiliation[d]{Department of Physics and Astronomy, University of Padova, via F. Marzolo 8, 35131 Padova, Italy}
\affiliation[e]{Marietta Blau Institute for Particle Physics, Austrian Academy of Sciences, Nikolsdorfer Gasse 18, Vienna, Austria}
\affiliation[f]{INFN Padova, via F. Marzolo 8, 35131 Padova, Italy}
\affiliation[g]{CNR-Istituto Officina dei Materiali (IOM), via A. Pascoli, Perugia, Italy}

\emailAdd{khaoula.aouadj@pg.infn.it}

\abstract{
Silicon carbide is a promising material for radiation-hard detectors due to its wide bandgap, low leakage current, high critical electric field, and high saturation velocity. A key obstacle for its use in high-radiation environments is the incomplete understanding of surface damage at the 4H-SiC/SiO$_2$ interface. In this work, we present a combined experimental and TCAD simulation study of X-ray radiation-induced surface damage on n-type 4H-SiC MOS capacitors fabricated at CNM (Centro Nacional de Microelectr\'{o}nica, Barcelona), irradiated up to \SI{10}{Mrad}. High-frequency (\SI{100}{\kilo\hertz}) and quasi-static capacitance-voltage (C--V) measurements are used to evaluate the evolution of fixed oxide charge density and interface trap density as a function of dose. A dose-dependent TCAD surface model is developed and validated against the full set of measurements. The optimised model reproduces the measured C--V characteristics across the entire irradiation range and provides a physically grounded baseline for simulations of n-type 4H-SiC-based detectors, and can be used as a starting point for predictive modelling of irradiated detectors.

}


\begin{document}
\maketitle
\flushbottom

\section{Introduction}
\label{sec:intro}
In the landscape of radiation-hard materials, 4H-SiC, a polytype of silicon carbide (SiC), has attracted increasing interest as an alternative to silicon for its ability to operate in extreme thermal and radiative regimes where conventional silicon reaches its functional limits. In High-Energy Physics (HEP) and nuclear environments, the primary appeal of 4H-SiC lies in its wide bandgap ($E_g \approx 3.26\,\mathrm{eV}$), which drastically reduces leakage current and enables room-temperature operation.

Over the last decade, although significant progress has been made in modelling radiation damage effects in 4H-SiC at the bulk level~\cite{gaggl2025}, the effect of irradiation on the 4H-SiC/SiO$_2$ interface remains a major research topic. The interface strongly affects the electrical behaviour of 4H-SiC‑based devices, and compared with the well-studied Si/SiO$_2$ system, the 4H-SiC/SiO$_2$ interface tends to host a higher density of intrinsic defects and a more complex distribution of interface states~\cite{fiorenza2019,fiorenza2020}, making both measurement and modelling more challenging. Therefore, a deeper understanding of these defects is needed to predict and improve the reliability of 4H-SiC-based devices under irradiation.

In this context, Technology Computer-Aided Design (TCAD) represents a powerful tool for the development of semiconductor devices. However, reliable simulations require accurate models, and no dose-dependent surface-damage model for 4H-SiC has yet been validated against a broad irradiation range in a single consistent framework. In particular, a quantitative description of radiation-induced defects as a function of dose is needed to interpret measurements and build numerical models capable of predicting 4H-SiC-based detector behaviour after irradiation.

This work presents a combined experimental and TCAD simulation study of X-ray radiation damage in n-type 4H-SiC MOS capacitors fabricated at CNM (Barcelona,  Spain). The devices were irradiated with X-rays in the dose range 50~krad--10~Mrad(SiO$_2$) and characterised using high-frequency (\SI{100}{\kilo\hertz}) and quasi-static C--V measurements. A dose-dependent TCAD surface model is developed based on experimental data, incorporating both fixed oxide charge and interface traps. The model is optimised using the full dose-dependent dataset and validated against the measurements, reproducing the observed C--V behaviour across the entire irradiation range.

The paper is organised as follows. Section~\ref{sec:devices} describes the devices, the experimental procedure, the C--V measurements, and the extracted parameters. Section~\ref{sec:tcad} outlines the TCAD modelling approach. Section~\ref{sec:conclusions} summarises the results and outlines possible next steps.


\section{Devices and Experiment}
\label{sec:devices}
The devices under study are n-type 4H-SiC MOS capacitors, fabricated at CNM (Centro Nacional de Microelectrónica, Barcelona). The gate dielectric consists of a thermally grown SiO$_2$ layer with a thickness of \SI{455}{\nano\meter}. The substrate is nitrogen-doped, with a donor concentration of $3 \times 10^{14}$~cm$^{-3}$. The samples were irradiated with X-rays at the Department of Physics and Astronomy of the University of Padua (Italy), at room temperature and without bias polarization, over five dose levels: \SI{50}{krad}, \SI{500}{krad}, \SI{1}{Mrad}, \SI{5}{Mrad}, and \SI{10}{Mrad}(SiO$_2$).

\begin{figure}[htbp]
  \centering
   \includegraphics[width=0.56\textwidth ,clip,trim=0 0 0 0]{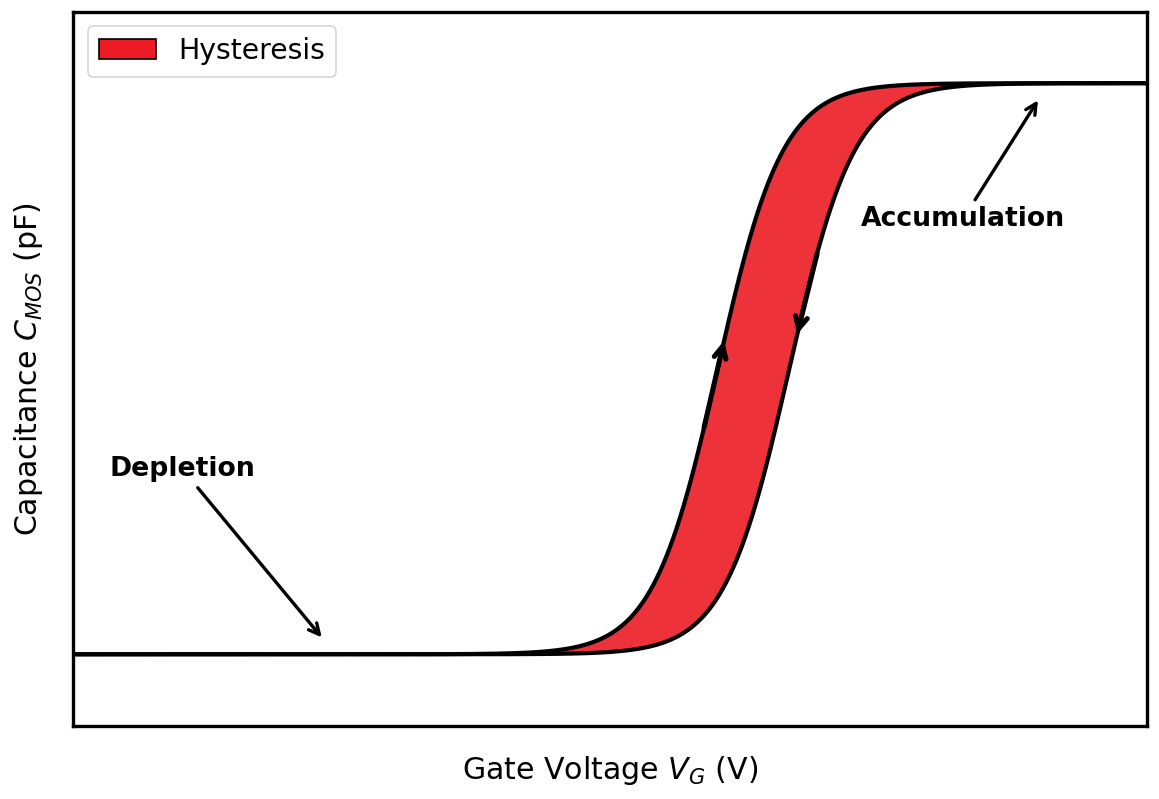}
  \caption{Representative C--V characteristics of an n-type MOS capacitor, showing both accumulation-to-depletion (AD) and depletion-to-accumulation (DA) voltage sweeps.}
  \label{fig:Hys}
\end{figure}

Capacitance-voltage (C--V) characteristics were measured at \SI {293}{\kelvin} using both high-frequency (HF) and quasi-static (QS) techniques. HF measurements were carried out at \SI{100}{\kilo\hertz}. At this frequency, interface traps do not respond to the AC signal and therefore do not contribute to the measured capacitance. As a result, the HF C--V curve mainly reflects the semiconductor's depletion response. QS measurements were performed using a voltage ramp of $0.24$\,V\,s$^{-1}$, slow enough to allow interface traps to follow the signal and contribute to the measured capacitance. For each dose, both accumulation-to-depletion (AD) and depletion-to-accumulation (DA) sweeps were recorded to evaluate hysteresis effects (see Figure~\ref{fig:Hys}). At least five independent measurements were performed on a minimum of two devices for each dose level to assess repeatability.

\begin{figure} [!htbp]
  \centering
  \begin{subfigure}[t]{0.49\textwidth}
    \centering
    \includegraphics[width=\textwidth,clip,trim=0 0 0 30]{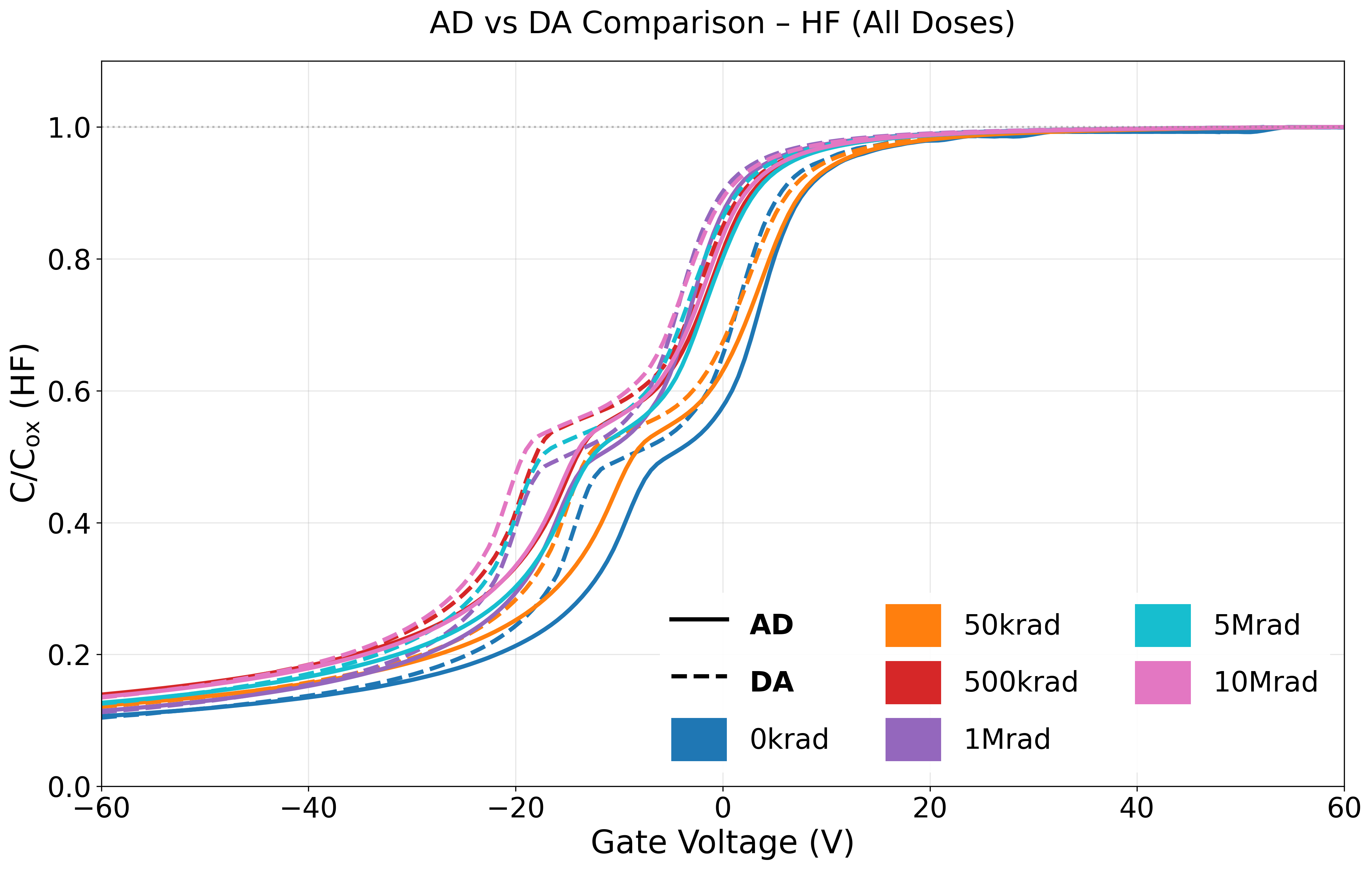}
  \end{subfigure}
  \hfill
  \begin{subfigure}[t]{0.49\textwidth}
    \centering
    \includegraphics[width=\textwidth , clip,trim=0 0 0 30]{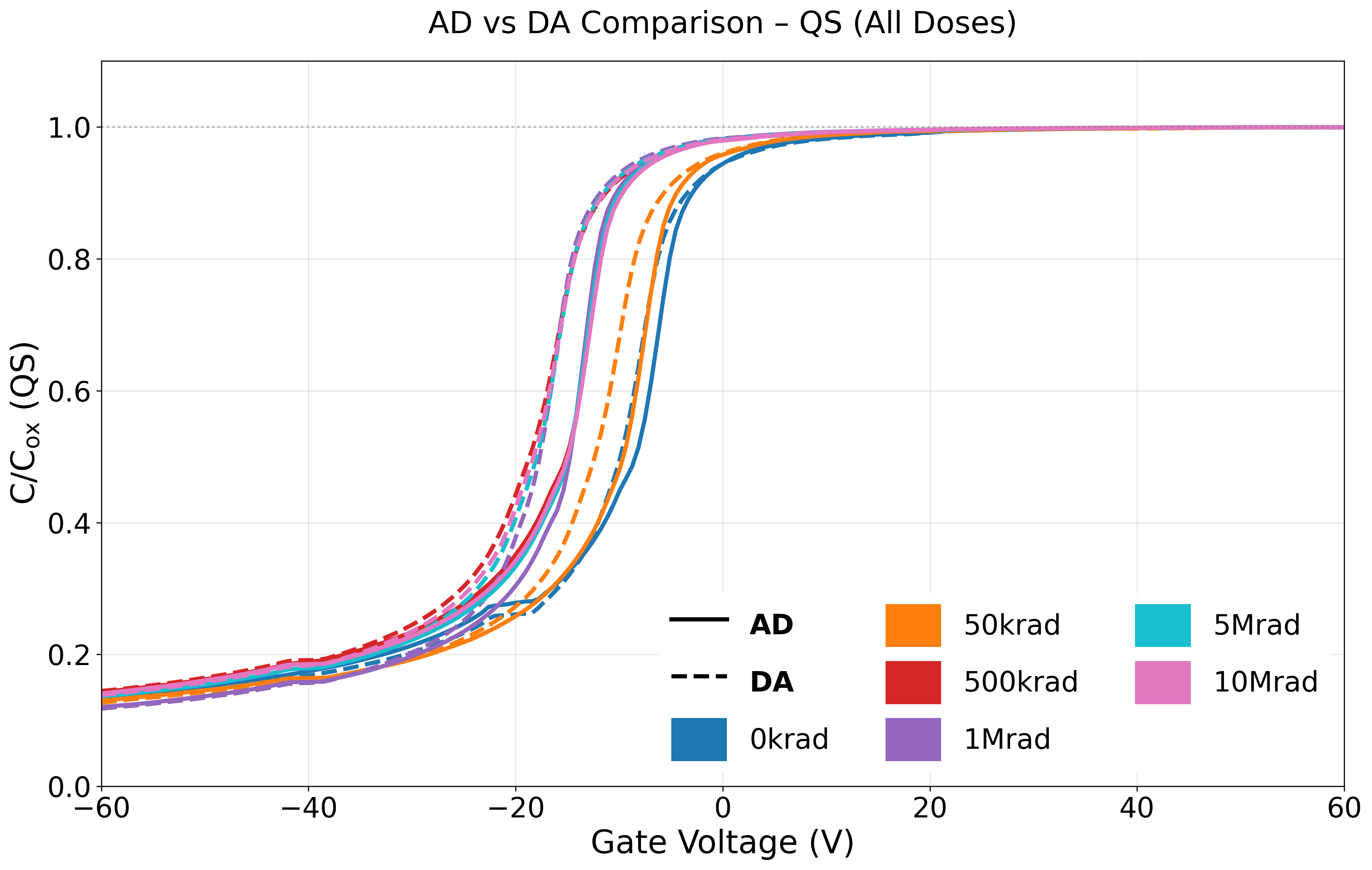}
  \end{subfigure}
  \caption{Normalised high-frequency (HF, left) and quasi-static (QS, right) C--V characteristics across all irradiation dose levels, for both AD (solid lines) and DA voltage sweeps (dashed lines). The curves show the average of the performed measurements.}
  \label{fig:HF_QS}
  
\end{figure}

As shown in Figure~\ref{fig:HF_QS}, both HF and QS C--V characteristics progressively shift toward more negative gate voltages with increasing radiation dose. The shift is monotonic from the pre-irradiation condition up to approximately \SI{500}{krad}, where the maximum displacement is observed. Beyond this dose, the curves exhibit a clear saturation behaviour, with the \SIrange{1}{10}{Mrad} characteristics remaining within experimental uncertainty for both AD and DA sweeps.
The observed negative shift of the C--V characteristics is consistent with the buildup of positive fixed oxide charge induced by irradiation.

The flat-band voltage $V_\text{FB}$ is extracted from the inflection point of the HF C--V curve, defined by the condition~\cite{10.1116/1.4802478}:

\begin{equation}
  \frac{d^2 C_\text{HF}}{dV_g^2}\bigg|_{V_g = V_\text{FB}} = 0
  \label{eq:vfb}
\end{equation}

\begin{figure}[htbp]
  \centering
   \includegraphics[width=0.60\textwidth ,clip,trim=0 0 0 35]{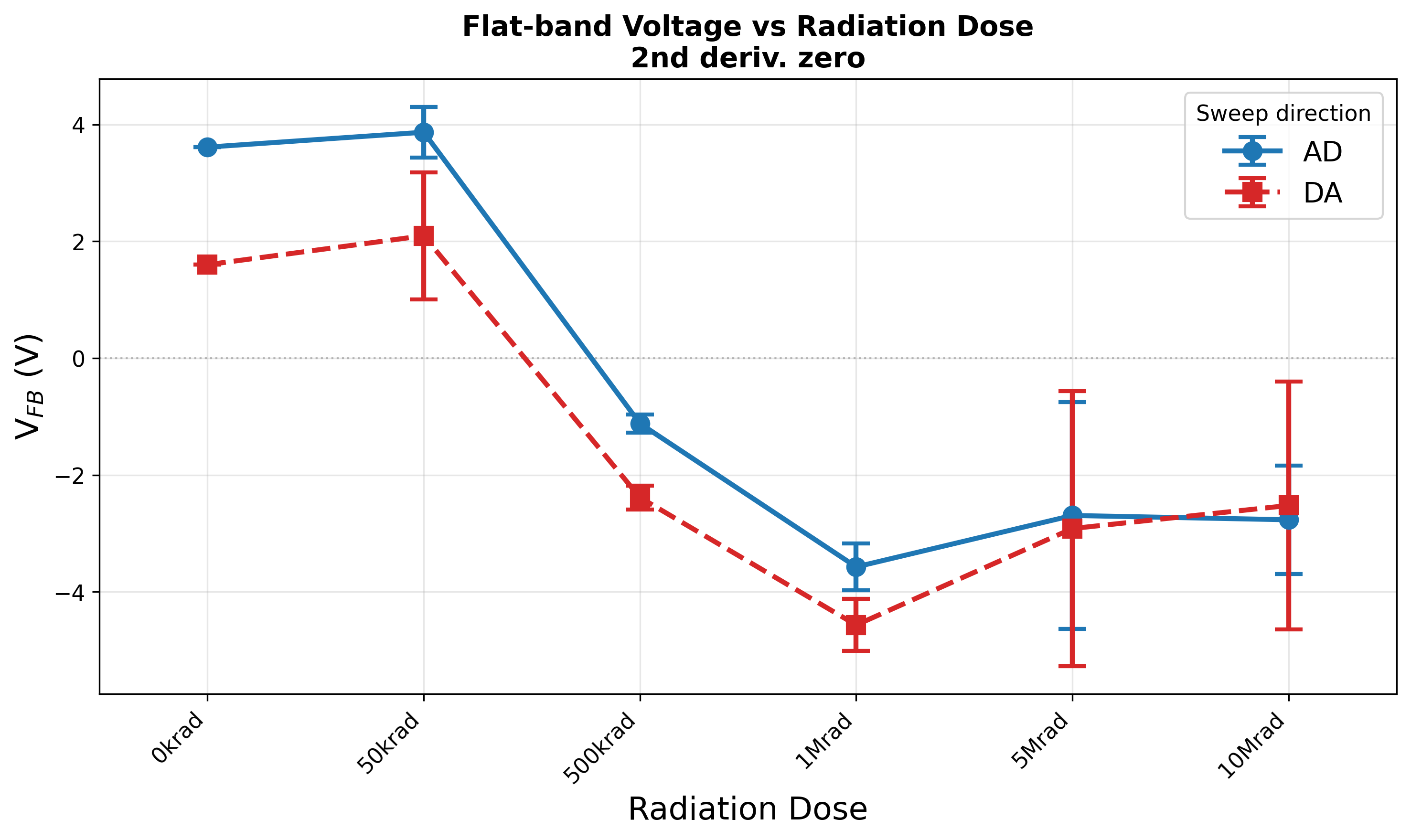}
  \caption{Flat-band voltage $V_\text{FB}$ as a function of absorbed dose, extracted from the inflection point of the HF C--V curve for both AD and DA sweeps. Error bars represent the standard deviation over different devices.}
  \label{fig:vfb}
\end{figure}

This approach provides a robust determination of $V_{\mathrm{FB}}$ even in the presence of high interface-state density and does not require knowledge of material or experimental parameters~\cite{10.1063/1.4982912}. 
The extracted $V_{\mathrm{FB}}$ is shown in Figure~\ref{fig:vfb} as a function of absorbed dose. It shifts toward more negative values up to approximately \SI{500}{krad}, and then saturates for both AD and DA, indicating an increase in oxide charge with dose that reaches saturation at higher doses.
The hysteresis $\Delta V = V_\mathrm{FB}^\mathrm{AD} - V_\mathrm{FB}^\mathrm{DA}$ decreases steadily from approximately \SI{2.0}{V} before irradiation to \SI{-0.24}{V} at \SI{10}{Mrad}, indicating a reduced contribution of slow border traps to the measured hysteresis with increasing radiation dose~\cite{nano10071332, FLEETWOOD2018266}.

\newpage
The interface trap density $D_\text{it}$ is estimated using the Hi-Lo method~\cite{nicollian1982}, which exploits the difference between QS and HF capacitances to determine $D_\text{it}$ across the 4H-SiC bandgap:

\begin{equation}
  D_\text{it}(E) = \frac{C_\text{ox}}{q\,A}\,
    \frac{C_\text{QS} - C_\text{HF}}{C_\text{HF}\!\left(C_\text{ox} - C_\text{QS}\right)}
  \label{eq:dit}
\end{equation}
where $q$ is the elementary charge, $A$ is the device area, $C_\text{ox}$ is the oxide capacitance, $C_\text{QS}$ is the quasi-static capacitance, and $C_\text{HF}$ is the high-frequency capacitance.

\begin{figure}[htbp]
  \centering
   \includegraphics[width=0.7\textwidth ,clip,trim=0 0 0 27]{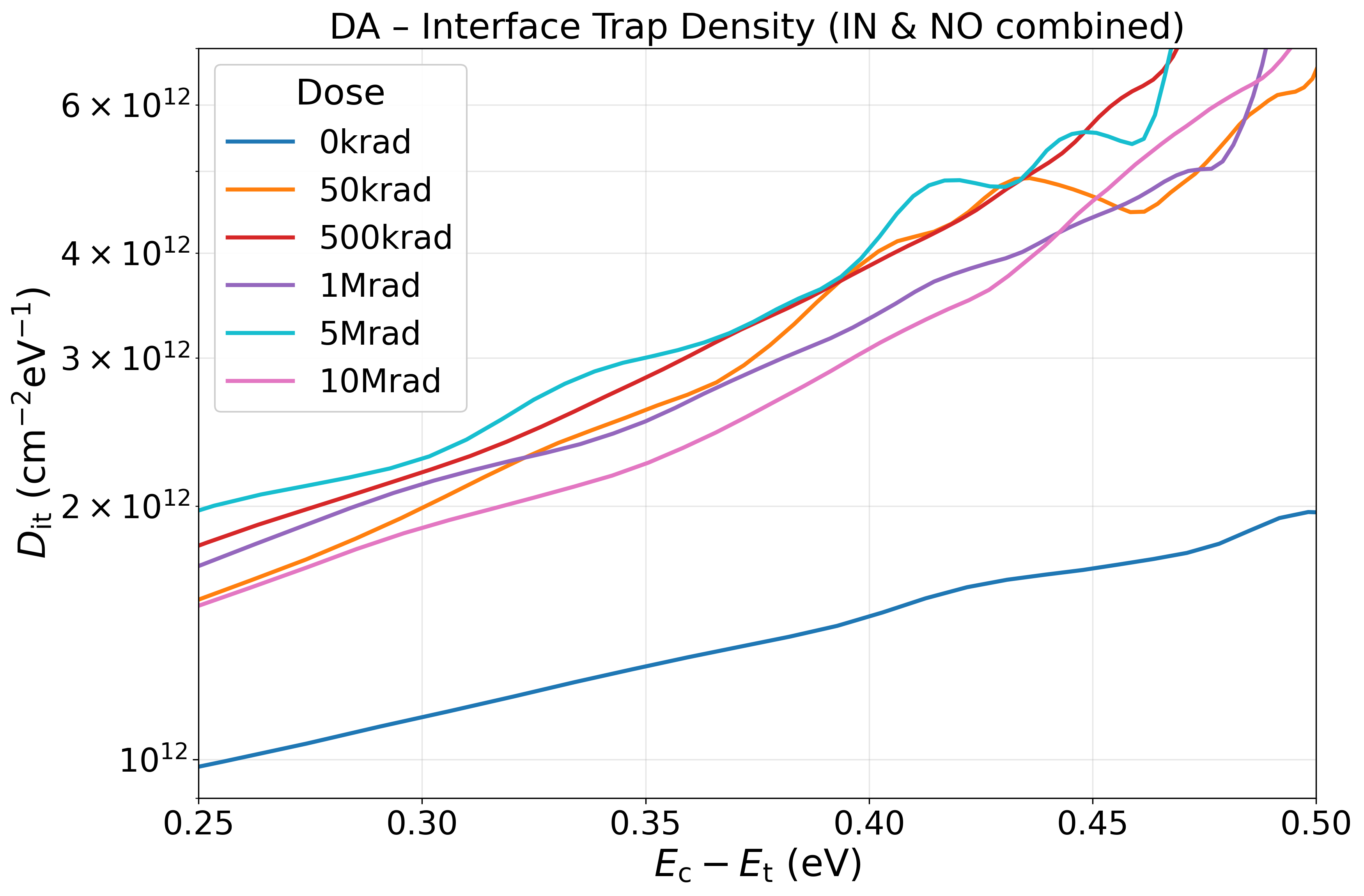}
  \caption{Interface trap density $D_{\mathrm{it}}$ as a function of energy below the conduction band edge, extracted using the Hi-Lo method for all dose levels.}
  \label{fig:Dit}
\end{figure}

The Hi-Lo method has known limitations for wide-bandgap materials~\cite{6422371}: traps near the band edges may exhibit emission times longer than the QS measurement timescale and therefore may not fully contribute to the measured capacitance, leading to an incomplete extraction of $D_{\mathrm{it}}$ in those energy regions. Consequently, the distributions shown in Figure~\ref{fig:Dit} are restricted to approximately $E_c-E_t = 0.20$--$0.50$\,eV below the conduction band edge, corresponding to the limited energy window effectively accessible under the measurement conditions ~\cite{6422371}.

To account for the interplay between different defect species, the experimental distributions $D_\text{it}$ served as baseline constraints for TCAD refinement, rather than static input parameters. This allowed for a more robust modeling of the device's electrical response.

\section{TCAD modelling approach}
\label{sec:tcad}

\subsection{Simulation setup}

The simulation domain is a two-dimensional MOS capacitor designed within the Synopsys\textsuperscript{\textregistered}~Sentaurus\textsuperscript{\texttrademark} TCAD environment. The actual geometry of the device has been properly taken into account by considering the Area Factor, which scales the 2D simulation results to match the physical dimensions of the tested devices. The substrate is uniformly doped n-type with a donor concentration of $3 \times 10^{14}$~cm$^{-3}$, consistent with the experimental device.

The simulations included a set of physical models appropriate for 4H-SiC MOS structures. Fermi–Dirac carrier statistics were enabled. Carrier transport was described using a doping-dependent mobility model including normal-field dependence. Incomplete ionisation of nitrogen donors was accounted for using a split-donor model. Recombination was modelled using Shockley–Read–Hall (SRH) statistics with doping and temperature dependence, while the intrinsic carrier concentration is evaluated using the Slotboom formulation ~\cite{BURIN2025112352}.

Both HF and QS C--V characteristics were simulated using transient voltage ramps reproducing the experimental sweep conditions, enabling direct comparison with the measurements.

\subsection{4H-SiC/SiO$_2$ Surface Model}

The 4H-SiC/SiO$_2$ surface is modeled using a combination of fixed oxide charge ($Q_{\text{ox}}$) and uniform distributions of interface traps ($D_{\text{it}}$). Specifically, the trap density is represented by two acceptor-like bands in the upper half of the bandgap and a single donor-like band in the lower half. The energy ranges and capture cross-sections ($\sigma_e$, $\sigma_h$) for these distributions are summarized in Table~\ref{tab:energ_all}.

\begin{table}[!htbp]
\centering
\caption{Interface defect model parameters: energy distributions and capture cross-sections for electrons ($\sigma_e$) and holes ($\sigma_h$).}
\label{tab:energ_all}
\smallskip
\begin{tabular}{lcccc}
\hline
Interface Defect & Energy Range [eV] & $\sigma_e$ [cm$^{2}$] & $\sigma_h$ [cm$^{2}$]\\
\hline
Acceptor Band 1  & $E_C - 0.815 < E_t < E_C$               & $1 \times 10^{-16}$ & $1 \times 10^{-15}$ \\
Acceptor Band 2  & $E_C - 1.63  < E_t < E_C - 0.815$       & $1 \times 10^{-18}$ & $1 \times 10^{-17}$ \\
Donor Band       & $E_V < E_t < E_V + 1.63$                & $1 \times 10^{-15}$ & $1 \times 10^{-16}$ \\
\hline
\end{tabular}
\end{table}

\begin{figure}[htbp]
  \centering
   \includegraphics[width=0.7\textwidth ,clip,trim=0 0 0 22]{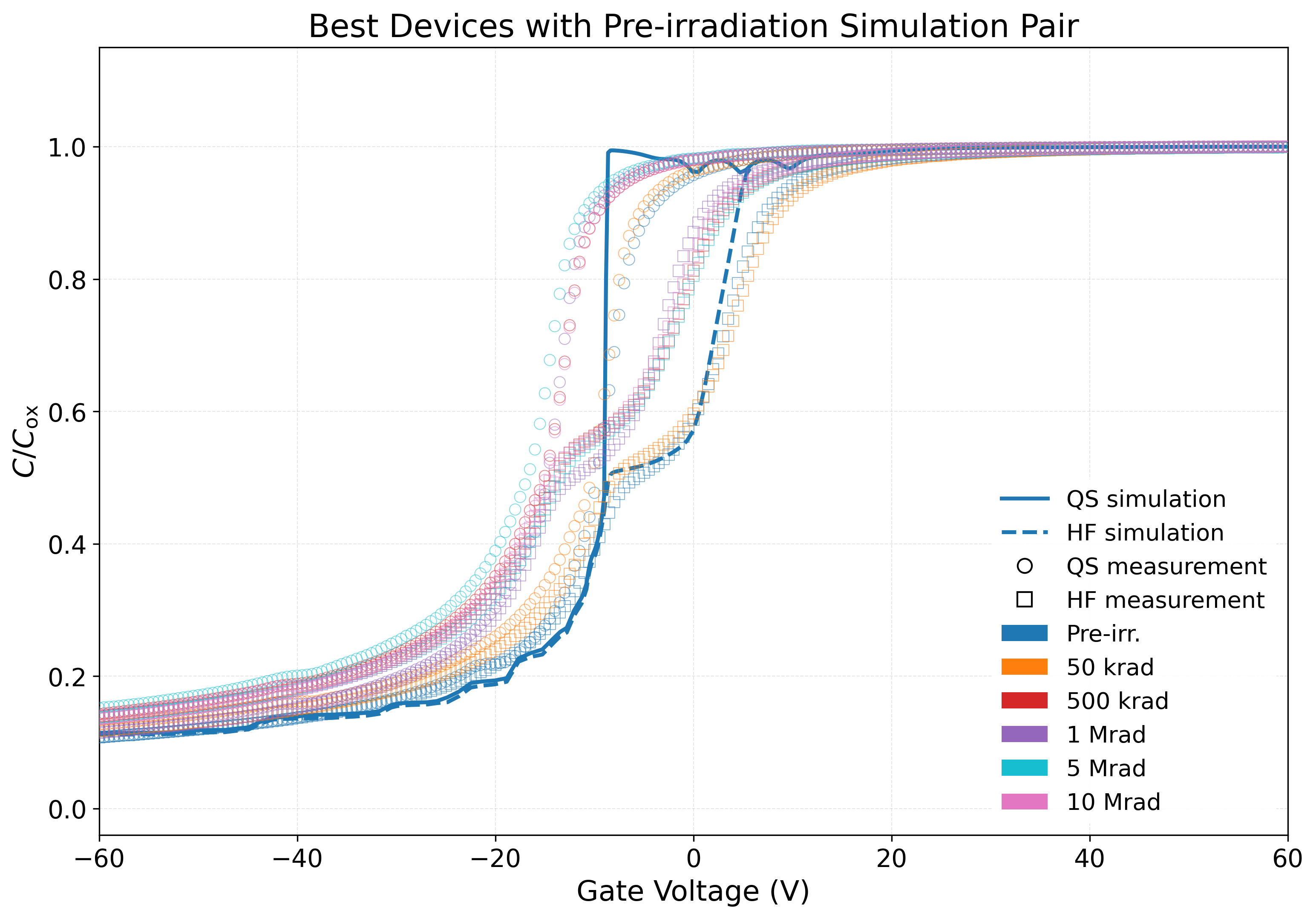}
  \caption{Comparison of simulated and measured HF and QS C--V characteristics using a single acceptor-band interface model.}
  \label{fig:failed}
\end{figure}

The initial modeling strategy followed a similar approach used for silicon-based detectors, employing one acceptor-like band extending from the conduction-band edge toward mid-gap and one donor-like band in the lower half of the bandgap \cite{Morozzi2021}. However, this simplified approach fails to reproduce the shape of the C–V characteristics in the transition from accumulation to depletion, in particular the “slow depletion” shoulder observed in both HF and QS measurements, as shown in Figure~\ref{fig:failed}.  In this configuration, the simulated capacitance drops too abruptly, indicating that a single uniform distribution cannot simultaneously describe traps with different response times.

To accurately capture this behavior, the acceptor-like distribution (spanning 1.63~eV below $E_C$) was split into two sub-bands of equal width. This dual-band approach provides the necessary flexibility to account for the coexistence of traps with different characteristic time constants, which is a known feature of the 4H-SiC/SiO$_2$ interface \cite{10.1063/1.2837028}. The shallower band is associated with larger capture cross-sections (faster states), whereas the deeper band is characterised by smaller capture cross-sections (slower states). This separation allows a more realistic description of the interface-state dynamics at the 4H-SiC/SiO$_2$ interface.

Within each band, the capture cross-sections are kept constant across all irradiation doses to ensure physical consistency of the model, while the corresponding interface-trap densities are independently computed at each dose point.
\newpage
\subsection{Simulation results}
The optimised interface parameters as a function of absorbed dose are summarised in Table~\ref{tab:params_all}. A progressive increase in both oxide charge and interface trap density is observed up to \SI{500}{krad}, followed by a clear saturation at higher doses consistent with the experimental saturation observed in $V_\text{FB}$.

The donor-like interface trap density has a weaker impact on the C–V characteristics of the investigated n-type devices due to the position of the Fermi level, which lies close to the conduction band. Based on Fermi–Dirac statistics, acceptor-like traps (negatively charged when occupied) are most effective when located below the Fermi level, while donor-like traps (positively charged when occupied) are effective when located above it. In n-type material, acceptor-like traps are predominantly occupied and contribute significantly to the interface charge, whereas donor-like traps remain mostly unoccupied and therefore electrically neutral. As a result, donor-like states are less active in charge exchange and have a limited influence on the measured capacitance. Consequently, their parameters are less strongly constrained by the measurements. Therefore, a simplified parameterisation is adopted in which the donor-like trap density is assumed equal to the density of the second acceptor-like band, primarily to ensure charge balance and numerical stability of the model.

\begin{table}[!htbp]
\centering
\caption{Optimised surface model parameters as a function of absorbed dose. Saturation is observed above \SI{500}{krad}.}
\label{tab:params_all}
\smallskip
\begin{tabular}{lcccc}
\hline
Dose & $Q_\text{ox}$ [cm$^{-2}$] & $D_\text{it}^\text{acc 1}$ [cm$^{-2}$eV$^{-1}$] & 
$D_\text{it}^\text{acc 2}$ [cm$^{-2}$eV$^{-1}$] & $D_\text{it}^\text{don}$ [cm$^{-2}$eV$^{-1}$] \\
\hline
Pre-irr.  & $1.2 \times 10^{12}$ & $1.8 \times 10^{12}$ & $1.0 \times 10^{12}$ & $1.0 \times 10^{12}$ \\
\SI{50}{krad}  & $1.5 \times 10^{12}$ & $2.2 \times 10^{12}$ & $1.2 \times 10^{12}$ & $1.2 \times 10^{12}$ \\
\SI{500}{krad} & $1.9 \times 10^{12}$ & $2.3 \times 10^{12}$ & $1.4 \times 10^{12}$ & $1.4 \times 10^{12}$ \\
\SI{1}{Mrad}   & $1.9 \times 10^{12}$ & $2.3 \times 10^{12}$ & $1.4 \times 10^{12}$ & $1.4 \times 10^{12}$ \\
\SI{5}{Mrad}   & $1.9 \times 10^{12}$ & $2.3 \times 10^{12}$ & $1.4 \times 10^{12}$ & $1.4 \times 10^{12}$ \\
\SI{10}{Mrad}  & $1.9 \times 10^{12}$ & $2.3 \times 10^{12}$ & $1.4 \times 10^{12}$ & $1.4 \times 10^{12}$ \\
\hline
\end{tabular}
\end{table}

\begin{figure}[!htbp]
  \centering
   \includegraphics[width=0.7\textwidth , clip,trim=0 0 0 45]{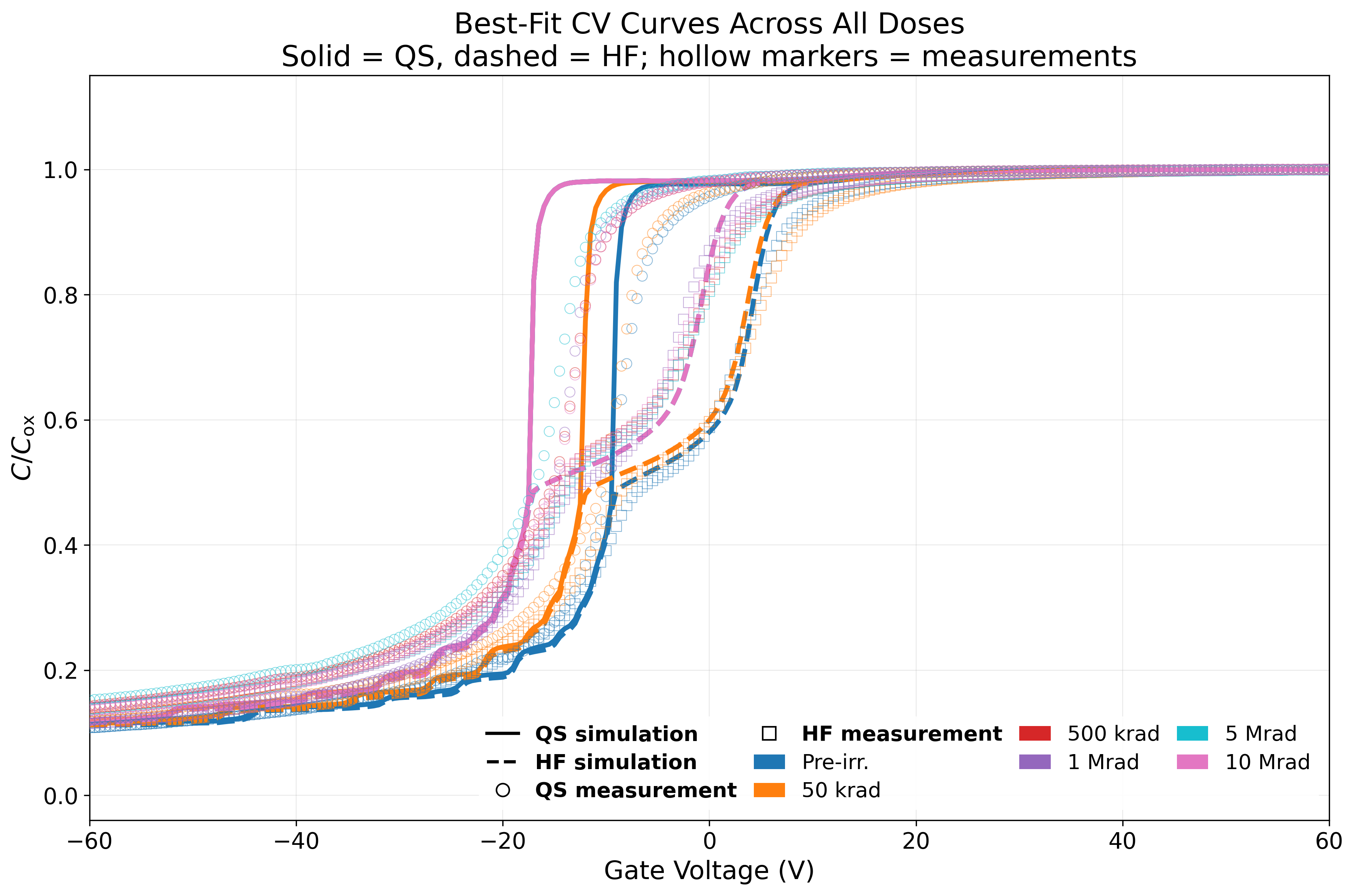}
  \caption{Comparison of simulated (lines) and measured (symbols) HF and QS C--V curves.}
  \label{fig:cv_comparison}
\end{figure}

Using these optimised parameters, as shown in Figure~\ref{fig:cv_comparison}, simulated curves show good agreement with the measurements over the full irradiation range, reproducing the overall shape and evolution of the C--V characteristics under both HF and QS conditions. In particular, the simulations correctly describe the transition from depletion to accumulation and the progressive shift of the curves toward negative gate voltages with increasing dose. The agreement is better for the HF curves, while moderate discrepancies are present in the QS depletion-to-accumulation transition region. Nevertheless, the model reproduces the main experimental features.

\newpage
\section{Conclusions and outlook}
\label{sec:conclusions}
In this work a combined experimental and TCAD simulation study of X-ray radiation damage at the 4H-SiC/SiO$_2$ interface in n-type MOS capacitors has been presented. Devices fabricated at CNM were irradiated up to 10~Mrad(SiO$_2$) and characterised using high-frequency and quasi-static C--V measurements, enabling the evaluation of oxide charge and interface trap density as a function of dose.

The extracted interface-trap density $D_\text{it}$ and flat-band voltage $V_\text{FB}$ exhibit a progressive evolution with irradiation dose, followed by a saturation above approximately \SI{500}{krad}. This behaviour is consistent with a saturation of radiation-induced defect generation at the interface and fixed charge in the oxide.

A dose-dependent surface model has been implemented in TCAD, including fixed oxide charge and interface states. Defect densities were refined to ensure consistency with experimental observations. Within this framework, a good agreement between simulated and measured C--V characteristics is obtained over the full dose range, accurately reproducing the capacitance behaviour from depletion to accumulation.

Overall, the proposed approach provides a consistent framework for the interpretation of radiation-induced surface effects in n-type 4H-SiC MOS structures. Future work will focus on applying this model to the simulation of 4H-SiC detectors, extending the study to proton irradiation and to devices fabricated with different processing technologies.

\acknowledgments
The authors acknowledge the support of INFN CSN5 through the SiC4GAIN research project. This work was also supported by the DRD3 project SiC-LGAD. The authors would like to thank the Institute of Microelectronics of Barcelona (IMB-CNM-CSIC) for supporting the production of the 4H-SiC detectors.
\bibliographystyle{JHEP}
\bibliography{biblio}
\end{document}